# First Demonstration of HZO/$\beta$-Ga$_2$O$_3$ Ferroelectric FinFET with Improved Memory Window

Seohyeon Park[1,†], Jaewook Yoo[1,†], Hyeojun Song[1], Hongseung Lee[1], Seongbin Lim[1], Soyeon Kim[1], Minah Park[1], Bongjoong Kim[4], Keun Heo[3], Peide D. Ye[2] and Hagyoul Bae[1,*]

[1]Electronic Engineering, Jeonbuk National University, Jeonju 54896, Republic of Korea

[2]School of Electrical and Computer Engineering, Purdue University, West Lafayette, Indiana 68588, USA

[3]Department of Semiconductor Science & Technology, Jeonbuk National University, Jeonju 54896, Republic of Korea.

[4]Department of Mechanical & System Design Engineering, Hongik University, Seoul 04066, Republic of Korea

[†]Seohyeon Park and Jaewook Yoo contributed equally to this work

*Author to whom correspondence should be addressed.

  Email address: hagyoul.bae@jbnu.ac.kr

**Abstract**

We have experimentally demonstrated the effectiveness of beta-gallium oxide ($\beta$-Ga$_2$O$_3$) ferroelectric fin field-effect transistors (Fe-FinFETs) for the first time. Atomic layer deposited (ALD) hafnium zirconium oxide (HZO) is used as the ferroelectric layer. The HZO/$\beta$-Ga$_2$O$_3$ Fe-FinFETs have wider counterclockwise hysteresis loops in the transfer characteristics than that of conventional planar FET, achieving record-high memory window (MW) of 13.9 V in a single HZO layer. When normalized to the actual channel width, FinFETs show an improved $I_{ON}/I_{OFF}$ ratio of 2.3×10$^7$ and a subthreshold swing value of 110 mV/dec. The enhanced characteristics are attributed to the low-interface state density ($D_{it}$), showing good interface properties between the $\beta$-Ga$_2$O$_3$ and HZO layer. The enhanced polarization due to larger electric fields across the entire ferroelectric layer in FinFETs is validated using Sentaurus TCAD. After 5×10$^6$ program/erase (PGM/ERS) cycles, the MW was maintained at 9.2 V, and the retention time was measured up to 3×10$^4$ s with low degradation. Therefore, the ultrawide bandgap (UWBG) Fe-FinFET was shown to be one of the promising candidates for high-density non-volatile memory devices.

# 1. Introduction

Monolithic $\beta$-Ga$_2$O$_3$ with an ultrawide bandgap (UWBG) of 4.6-4.9 eV has been identified as an emerging candidate for next-generation electronic devices. Its UWBG enables the $\beta$-Ga$_2$O$_3$ material to have high breakdown electric fields, high electron mobility, sustainability under large electric fields, high power, and high operating temperatures [1]-[5]. These properties make $\beta$-Ga$_2$O$_3$ particularly attractive for electronic devices that exhibit high voltage operations in harsh environments [6]-[8].

Meanwhile, hafnium zirconium oxide (HZO) has shown ferroelectric behaviors at high temperatures. HZO layer is a promising component given its compatibility with complementary metal-oxide-semiconductor (CMOS) in back-end-of-line (BEOL) thermal budget (<400 °C) and strong ferroelectricity even at very thin thickness (< 10 nm) [9]-[12]. Due to these characteristics, HZO can be applied to various devices, including ferroelectric field-effect transistor (FeFET), ferroelectric random-access memory (FeRAM), negative capacitance field effect transistor (NCFET), and ferroelectric tunnel junction memory (FTJ) [13], [14]. FeFETs have been highlighted for their ability to be fabricated by inserting a ferroelectric layer into the gate stack of a conventional MOSFET, allowing it to exhibit non-volatile memory behaviors while maintaining the same footprint as conventional FETs. Ferroelectric UWBG semiconductor devices have the potential to fill the need for robust electrical operation of neuromorphic applications. A HZO/$\beta$-Ga$_2$O$_3$ ferroelectric FET recently achieved 94% on-chip learning accuracy at high temperatures, with improved electrical performance [15]. From the perspective of channel structures, the 3-dimensional architecture – providing both scalability and performance enhancement – is showing significantly improved results as presented in previous studies.

In this work, HZO/$\beta$-Ga$_2$O$_3$ ferroelectric FinFET (Fe-FinFET) with a fin width of 50 nm was fabricated. The fabricated Fe-FinFET presented robust electrical performances with faster electric dipole switching, high $I_{ON}/I_{OFF}$ ratio, steep subthreshold swing (SS), large memory window (MW), and low interface state density ($D_{it}$) compared with planar structured FeFET. The Fe-FinFET also exhibited improved performances in retention time and endurance, making it a significant candidate for memory devices. Our experimental results indicate that the HZO/$β$-Ga$_2$O$_3$ Fe-FinFET is promising for highly reliable and scalable devices due to the remarkably improved gate controllability.

## 2. Device fabrication

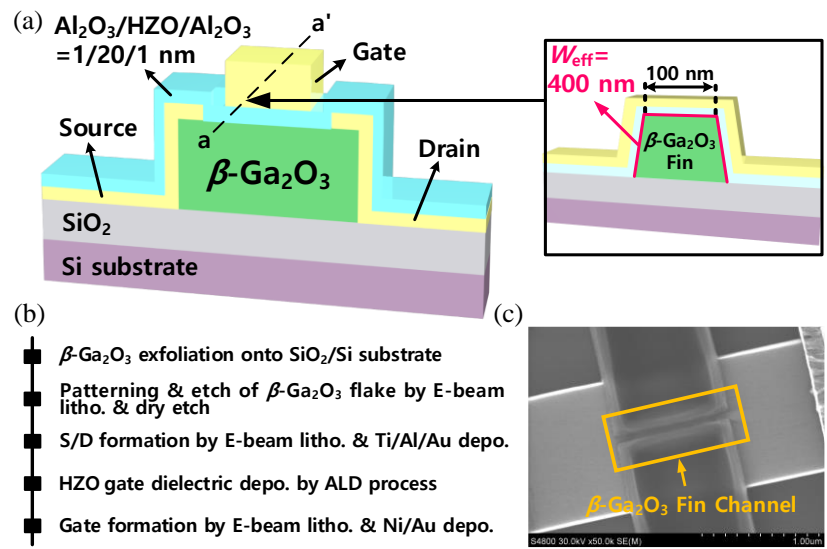

Fig. 1. (a) Schematic of HZO/$β$-Ga$_2$O$_3$ ferroelectric FinFET. (b) Key process steps in device fabrication. (c) SEM image of the fabricated 3D fin structure after RIE etching process.

Fig. 1 (a) shows the cross-sectional view of the fabricated device along both channel width and length directions and Fig. 1 (b) lists the key fabrication steps. For the HZO/$β$-Ga$_2$O$_3$ Fe-FinFET, thin nanomembrane flakes were realized by cleaving the $β$-Ga$_2$O$_3$ bulk substrate into small pieces and employing a mechanical exfoliation technique with adhesive tape [16]. The exfoliated flake with a thickness of 130 nm was transferred onto 270 nm $SiO_2$/p+ type Si substrate.

The active region with a fin width of 100 nm was defined by electron beam lithography (EBL) and dry etch process. Source and drain electrodes were patterned by EBL. Fig. 1 (c) presents the scanning electron microscope (SEM) image of the etched β-$Ga_2O_3$ channel with a width of 100 nm. Before metallization, an Ar plasma bombardment with a radio frequency power of 100 W was applied to improve the contact resistance ($R_C$) by generating oxygen vacancies to enhance the surface of β-$Ga_2O_3$ flakes. Ti/Al/Au (15/60/50 nm) metal electrodes were deposited using electron beam evaporation (EBE) and the lift-off process. A ferroelectric HZO gate stack of $Al_2O_3$/HZO/$Al_2O_3$ (1/20/1 nm) was deposited by the atomic layer deposition (ALD) process. Applying the 1 nm amorphous $Al_2O_3$ layer on the bottom of the stack achieved better interface quality, and the top $Al_2O_3$ was deposited to avoid degradation of the HZO by reaction with other contamination sources. EBL was carried out for gate patterning, and Ni/Au (50/50 nm) gate metal was formed via EBE and lift-off process. The device was suitably fabricated with each part of the β-$Ga_2O_3$ flake and HZO layer. Their components and accurate locations were confirmed through the STEM-EDS mapping images, as shown in Fig. 2.

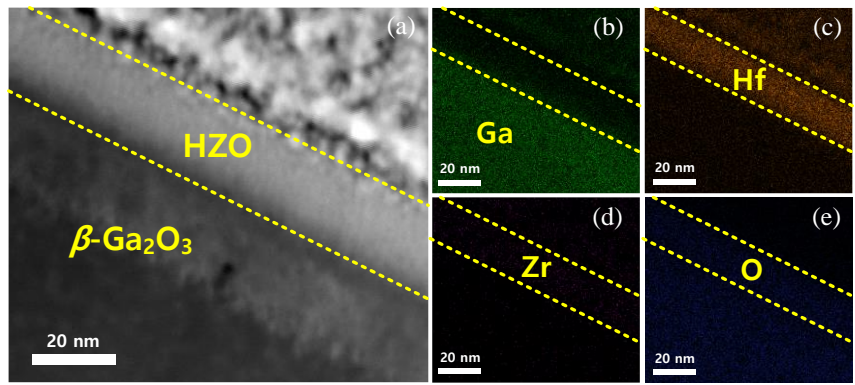

Figure 2. (a) TEM image of the fabricated HZO/β-$Ga_2O_3$ layers and EDS elemental mapping with scanning TEM images of (b) Ga, (c) Hf, (d) Zr, and (e) O.

## 3. Results and discussion

### 3.1 Electrical analysis

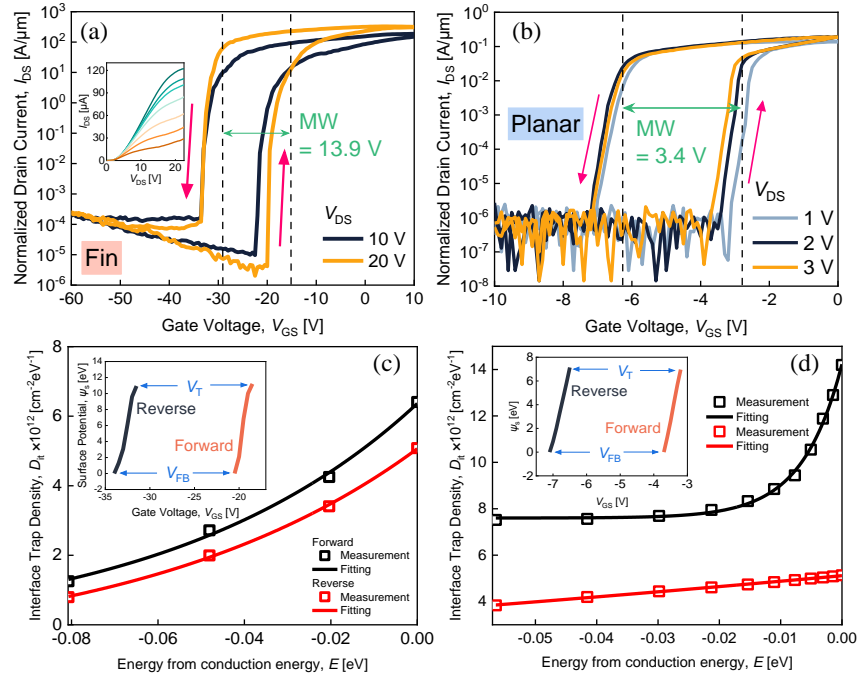

Figure 3. (a) $I_{DS}$-$V_{GS}$ with counter-clockwise hysteresis loop of (a) FinFET and $I_{DS}$-$V_{DS}$ (inset) and (b) Planar FET. $D_{it}$ and surface potential (inset) for Fin and Planar structure, shown in (c) and (d), respectively.

The drain current ($I_{DS}$)–gate-to-source voltage ($V_{GS}$) curves of FinFET and planar FET are shown in Fig. 3 (a) and (b), which considered the normalized to actual channel width. Typical counter-clockwise hysteresis loops with good polarization switching behaviors were observed in the transfer curves of both FETs. The key device performance parameters of the FinFET are more remarkable than those of the planar FET. With FinFET structure, these properties can be improved due to enhanced gate controllability with surrounding gate architecture and the strengthening electric field across the HZO layer. The remarkably broad $V_{GS}$ operation of Fe-FinFET is due to the UWBG of $\beta$-$Ga_2O_3$, which magnifies the capability of switching high voltages. Indeed, the FinFET shows a record-high MW of 13.9 V at the maximum sweeping voltage from -60 V to 10 V on the HZO single layer, which is over 4.2 times larger than that of 3.3 V for the planar FET at the linear region, where the MW was extracted from the threshold voltage ($V_T$) difference. This signifies that wide MW is required for practical memory applications for stand-alone data storage to separate the program and erase operation [17], [18]. By obtaining a large MW, moreover, the ferroelectric memory can be maintained with a capacity for multiple-level memory states for a long duration as future robust neuromorphic devices [15]. In this way, the MW of 13.9 V demonstrates the strength of the $\beta$-Ga$_2$O$_3$ Fe-FinFET for being able to be used in oxide-based non-volatile memory. We extracted the surface potential ($\psi_S$) and distribution of $D_{it}$ in both forward and reverse directions of Fe-FinFET and Planar FeFET as shown in Fig. 3 (c) and (d), respectively [2], [19]. The Fe-FinFET exhibits a larger $\psi_S$ value within the same voltage range, indicating that the HZO is subjected to a higher electric field and possesses better-switching characteristics. The $D_{it}$ for forward/reverse direction of Fe-FinFET is found to be suppressed down to lower than $6.4\times10^{12}$ cm$^{-2}$eV$^{-1}$, which is compared to Planar FeFET results at $1.4\times10^{13}$ cm$^{-2}$eV$^{-1}$.

Table I
COMPARISON OF ELECTRICAL PERFORMANCES
OF FIN AND PLANAR STRUCTURES

|  | Fe-FinFET | Planar FeFET |
|---|---|---|
| $I_{ON}$ [A/μm] | 16.93 | $3\times10^{-3}$ |
| $I_{ON}/I_{OFF}$ ratio | $2.3\times10^{7}$ | $8.1\times10^{4}$ |
| Subthreshold swing, SS [mV/dec] | 110 | 160 |
| Memory Window, MW [V] | 13.9 | 3.3 |
| $D_{it}$ [cm$^{-2}$eV$^{-1}$] | ~ $6.4\times10^{12}$ | ~ $1.4\times10^{13}$ |

Table I presents a comparison of the properties of the HZO/$\beta$-Ga$_2$O$_3$ Fe-FinFET with planar-structured FeFET. Compared with the Planar FeFET, the Fe-FinFET in this study exhibited enhanced electrical performances and excellent memory characteristics under high density.

## 3.2 TCAD simulation

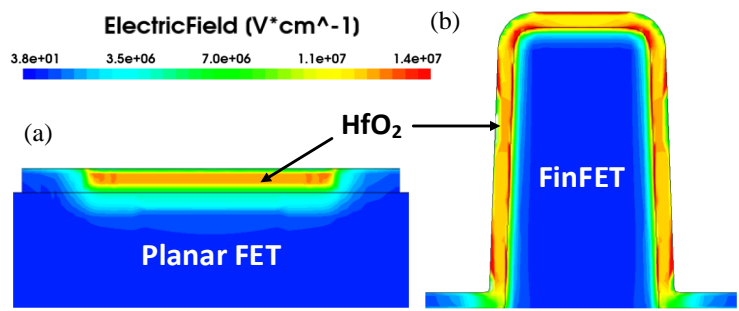

Figure 4. TCAD simulation results for the electric field at the HfO$_2$ layer of (a) planar FET and (b) FinFET.

For the ferroelectric material to fully polarize, a sufficient electric field across the ferroelectric layer is necessary [20]. The electric field enhancement in the conventional FinFET structure of Si substrate and HfO$_2$ gate insulator was understood using technology computer-aided design (TCAD) simulation (Synopsys Sentaurus) [21]. The simulation condition was $V_G$=5 V, $V_D$ = 0 V, $V_S$ = 0 V, and HfO$_2$ layer thickness=2 nm. As depicted in Fig. 4, the FinFET has a higher electric field across the HfO$_2$ layer than planar-structured FET; it is confirmed that the polarization is enhanced by the larger electric field across the ferroelectric layer, resulting in wider MW [22].

## 3.3 Reliability characteristics

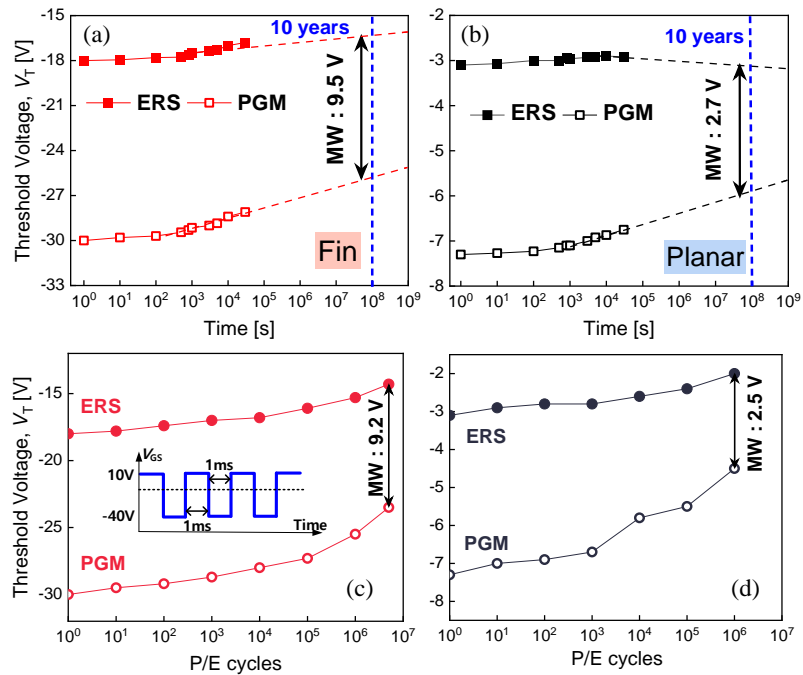

Fig. 5. Retention characteristics for (a) Fe-FinFET and (b) Planar FeFET. Endurance characteristics for (c) Fe-FinFET and (d) Planar FeFET.

Retention and endurance characteristics for investigating the ferroelectric memory reliability of the HZO/$\beta$-Ga$_2$O$_3$ Fe-FinFET are presented in Fig. 5. According to the retention characteristics, when the linear extrapolated to 10 years, the MW of the FinFET remains over 9.5 V. The endurance performance as a memory device was tested by program/erase (PGM/ERS) cycling with a fixed pulse width of 1 ms and repeated voltages of -40 V and 10 V. The Fe-FinFET shows lower deterioration up to $10^6$ cycles or more in the larger MW, compared to the Planar FeFET.

## 4. Conclusion

This study presented the first HZO/$\beta$-Ga$_2$O$_3$ Fe-FinFET approach with 50 nm of fin width. We report that the UWBG Fe-FinFET achieves a record-high MW of 13.9 V in a single ferroelectric layer. Additionally, FinFET structure exhibited improved SS of 110 mV/dec, high $I_{on}/I_{off}$ ratio of 3.8×10$^6$, low $D_{it}$, long retention of 10 years, and high endurance of 10$^6$ cycles under scaled dimensions. The enhanced polarization by a large electric field across the ferroelectric layer of FinFET was verified using TCAD simulation. This work demonstrated that the approach enables the HZO/$\beta$-Ga$_2$O$_3$ Fe-FinFET to be both highly reliable and scalable, and be able to be used in applications with future high-performance memory devices.

**Acknowledgment**

This work was supported by the National Research Foundation of Korea (NRF) grant funded by the Korea government (MSIT) (2022R1F1A1071914) and in part by the Nano-Material Technology Development Program (2009-0082580) through the National Research Foundation (NRF) of Korea funded by the Ministry of Science. The EDA tool was supported in part by the IC Design Education Center (IDEC).